\documentclass[sigconf, nonacm]{acmart}

\usepackage{algorithm}
\usepackage{algpseudocode}
\usepackage{amsmath}
\usepackage{subcaption}
\usepackage{placeins}
\usepackage{graphicx}
\usepackage{makecell}
\usepackage{multirow}
\usepackage{enumitem}

\graphicspath{{figures/}}

\begin{document}
\title{The Invocation Bottleneck in LLM-Based Lossless Compression: What Can Be Removed and What Must Be Traded}

\settopmatter{authorsperrow=4}

\author{Zhe Yan}
\affiliation{%
  \institution{Xidian University}
  \country{China}
}
\email{Angie.zheyan@gmail.com}

\author{Yihao Liu}
\affiliation{%
  \institution{Tsinghua University}
  \country{China}
}
\email{liuyihao24@\\mails.tsinghua.edu.cn}

\author{Antonio Boffa}
\affiliation{%
  \institution{EPFL}
  \country{Switzerland}
}
\email{antonio.boffa@epfl.ch}

\author{Huanchen Zhang}
\authornote{Also affiliated with the Shanghai Qi Zhi Institute. Corresponding author.}
\affiliation{%
  \institution{Tsinghua University}
  \country{China}
}
\email{huanchen@tsinghua.edu.cn}

\begin{abstract}
Large language models (LLMs) compress text far better than classical compressors, but LLM-based compressors are slow, particularly during decompression. We argue that their decoding time has two fundamentally different sources. Execution redundancy, such as per-token host transfers, CPU-side coding, and token-by-token encoding, is work that the coded output does not require and can be removed without changing a single output bit. Sequential dependency, one model call per decoded token, is inherent to lossless decoding and cannot be removed by better execution, only traded for more calls or for more bits. We remove execution redundancy with a GPU-resident pipeline whose block-parallel encoding is made lossless in half precision by batch-invariant kernels, speeding up arithmetic coding by up to 5.8$\times$ in encoding and 3.0$\times$ in decoding without changing its compression rate. We then study the two ways of trading sequential dependency, speculative dual-model coding (SDMC) and multi-token single-model coding (MSMC), characterize the compression cost of predicting several tokens from a shared prefix, and identify the workloads in which fewer model calls are worth that cost.
\end{abstract}

\maketitle
\noindent\textbf{Artifact Availability:} The source code, data, and other artifacts are available at \url{https://github.com/Angie-Yan/LLM-based-Data-Compression}.

\section{Introduction}
\label{sec:intro}

Data compression is fundamentally a problem of density estimation paired with entropy coding~\cite{shannon1948mathematical}: the more accurately a probabilistic model predicts upcoming symbols, the closer the resulting bitstream approaches the entropy of the source. Early learned compressors relied on lightweight statistical models or compact recurrent networks~\cite{knoll_cmix, bellard2019nncp}, whose limited modeling capacity kept their compression rates well above the entropy of natural language and structured text. Large language models (LLMs) change this picture: as highly expressive next-token predictors, they make the long-standing observation that language modeling is compression~\cite{deletang2023language} directly usable in practice.

LLM-based compression is increasingly relevant to long-lived archival storage, including source-code repositories, historical data-lake partitions, high-density system logs, and scientific datasets. These environments retain data for extended periods while still requiring reasonably fast access. In these settings, the one-time compute cost of compression, and of adapting auxiliary components to a data domain, can be amortized over a long retention period and over all data of that domain. Furthermore, modern data platforms increasingly use GPUs for both analytical processing and AI inference~\cite{jiang2026gust}. Where such accelerators and models are already deployed, reusing them for storage operations such as decompression lowers the marginal cost of LLM-based compression, turning it from a dedicated compute expense into a plausible use of existing capacity.

LLM-based compressors pair the model's conditional probabilities $P(t_i \mid t_{<i})$ with an entropy coder, most commonly Arithmetic Coding (AC)~\cite{rissanen1976generalized,witten1987arithmetic}, as in~\cite{valmeekam2023llmzip,finezip,deletang2023language}, and reach compression rates well beyond traditional compressors such as gzip~\cite{deutsch1996gzip}, zstd~\cite{collet2021zstd} and xz~\cite{collin_xz}, the strongest of which in our evaluation is xz. Their weakness is speed. A recent economic analysis~\cite{waiting2026cidr} concludes that, at current prices, their compute cost takes years to amortize, and follow-up work~\cite{adams2026highthroughput} shows that in the compression direction, arithmetic coding rather than model inference is the main bottleneck. Decompression is harder to accelerate, because each token's distribution depends on the tokens decoded before it. Our profiling of the public implementation accompanying~\cite{waiting2026cidr} shows that here, too, inference is not the only cost: on our hardware, it decodes Wikipedia text at 14.3\,KB/s, and CPU-side arithmetic coding together with copying each probability vector to the host accounts for about two thirds of decode time.

We argue that decoding time in LLM-based compression has two fundamentally different sources, which call for different treatment. \emph{Execution redundancy} is work that the coded output does not require: transferring a full probability vector to the host for every token, building cumulative distributions on the CPU, and encoding token by token although the encoder knows the whole input. It can be removed without changing a single output bit. \emph{Sequential dependency} is inherent to lossless decoding: the decoder cannot evaluate a token's distribution before it has recovered the preceding tokens. With a fixed model, it cannot be removed by better execution, only traded. Sequential calls can be made cheaper for some tokens by delegating them to a second and smaller model, although every token still requires one sequential call. Alternatively, calls can be made fewer by coding several tokens from a shared prefix; for coders that predict each offset independently, this costs bits, only part of which better predictors can recover.

\begin{figure}[t]
\centering
\includegraphics[width=\columnwidth]{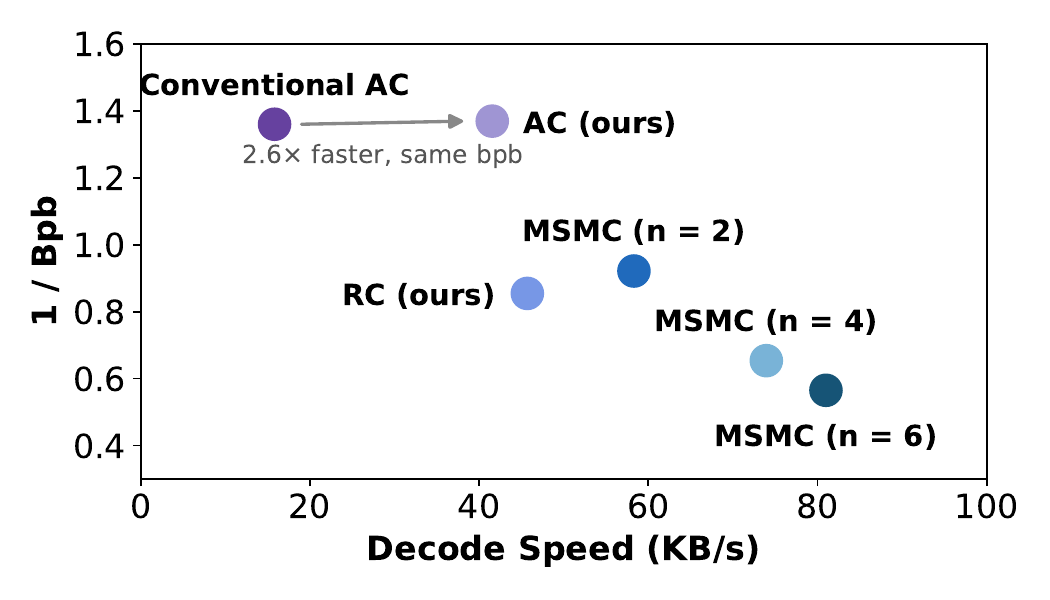}
\caption{Compression rate (as 1/bpb, higher is better) versus decode throughput, averaged over Wiki and Code (Qwen2.5-0.5B~\cite{qwen2025qwen25}, FP16, $B{=}256$). Removing execution redundancy speeds up arithmetic coding without changing bpb; resolving more tokens per call (MSMC) trades bpb for further decode throughput.}
\label{fig:pareto-intro}
\end{figure}

We study both limits on a common backbone, the same evaluation files, and a byte-exact losslessness requirement. For execution redundancy, we build a GPU-resident coding pipeline that encodes whole blocks by teacher forcing and keeps vocabulary-sized work on the GPU, and we make it lossless in half precision with batch-invariant kernels~\cite{he2025nondeterminism}, which give encoder and decoder bit-identical predictions although they call the model with different shapes. We pair the pipeline with arithmetic coding and with Rank-Tier Coding, a simple fixed-length rank code in the spirit of prior rank-based schemes~\cite{shannon1951prediction, valmeekam2023llmzip}, which we use as a low-cost reference point to measure how much the choice of coder still matters. For sequential dependency, we evaluate one representative design for each way of paying, as a probe of the trade-off rather than a replacement for token-by-token decoding: speculative dual-model coding (SDMC), which keeps full conditioning and pays in calls, and multi-token single-model coding (MSMC), which reduces calls and pays in bits. Our individual components build on existing techniques, namely batch-invariant kernels~\cite{he2025nondeterminism}, rank-based coding, speculative decoding~\cite{leviathan2023speculative}, and multi-token heads~\cite{cai2024medusa}; the contribution lies in making them lossless and composable in a compression pipeline and in characterizing what each gains and costs. Figure~\ref{fig:pareto-intro} summarizes the resulting operating points.

Our contributions are as follows. First, we separate decoding time into execution redundancy, which can be removed without changing the output, and sequential dependency, which can only be traded for calls or for bits, and show where each dominates (Section~\ref{sec:background}). Second, we apply batch-invariant kernels~\cite{he2025nondeterminism} to make block-parallel encoding lossless in half precision, and build a GPU-resident pipeline that removes redundancy from decoding as well. With the same model and probabilities, arithmetic coding becomes 4.4--5.8$\times$ faster for encoding and 2.3--3.0$\times$ faster for decoding; entropy coding then takes under 3\% of decode time, so any faster coder could add at most about 10\% (Sections~\ref{sec:substrate} and~\ref{subsec:time}). Third, for multi-token coders that predict each offset independently from a shared prefix, we identify a code-length overhead set by the information lost between offsets, and show that per-offset code lengths account for the measured compression loss of 2--3$\times$ at $n{=}6$ (Sections~\ref{subsec:price} and~\ref{subsec:why-degrade}). Keeping full conditioning with a draft model instead saves little decode time and halves encode throughput (Section~\ref{subsec:sdmc-vs-msmc}). Finally, we model and validate both sides of the multi-token trade-off: fewer calls pay off when decoding is latency-bound, with a 5.7$\times$ speedup for single-stream decoding, but much less at large batch sizes, where batching already fills the GPU (Section~\ref{subsec:when-pays}).
The best LLM-based operating points produce output 2.4--2.7$\times$ smaller than xz-9, while remaining far slower. We release our implementation, including the batch-invariant kernels.

Section~\ref{sec:background} analyzes where decoding time goes and states the two limits. Section~\ref{sec:substrate} removes execution redundancy. Sections~\ref{sec:speculative} and~\ref{sec:msmc} study the two ways of trading sequential dependency, SDMC and MSMC. Section~\ref{sec:experiments} evaluates them, Section~\ref{sec:related} discusses related work, and Section~\ref{sec:discussion} concludes the lessons and open directions.
\section{Two Limits on Decoding Throughput}
\label{sec:background}
\label{sec:AC}
 
\subsection{LLM-Based Compression and Where Its Time Goes}
\label{subsec:time-goes}
 
\paragraph{Framework.}
Standard LLM-based lossless compression combines an autoregressive model with an entropy coder, commonly Arithmetic Coding (AC)~\cite{rissanen1976generalized, witten1987arithmetic}, as in~\cite{valmeekam2023llmzip, deletang2023language, finezip}. Given a token sequence $(x_1, \dots, x_N)$, the model estimates $P(x_i \mid x_{<i})$ at each position, and by Shannon's source coding theorem~\cite{shannon1948mathematical} the ideal code length is
\begin{equation}
L \approx -\sum_{i=1}^N \log_2 P(x_i \mid x_{<i}),
\label{eq:codelength}
\end{equation}
which arithmetic coding approaches within a few bits. The model therefore determines the compression rate, and how the model and the coder are executed determines the speed. Encoding and decoding differ in one essential respect. The encoder knows the whole sequence and can in principle evaluate all positions in parallel; the decoder cannot, because the distribution at position $i$ depends on tokens it has not yet recovered. Practical systems therefore split the input into $B$ independent streams and advance them in lockstep, so that each model call serves $B$ tokens~\cite{waiting2026cidr}, while decoding within each stream remains sequential.
 
\begin{figure}[t]
\centering
\includegraphics[width=\columnwidth]{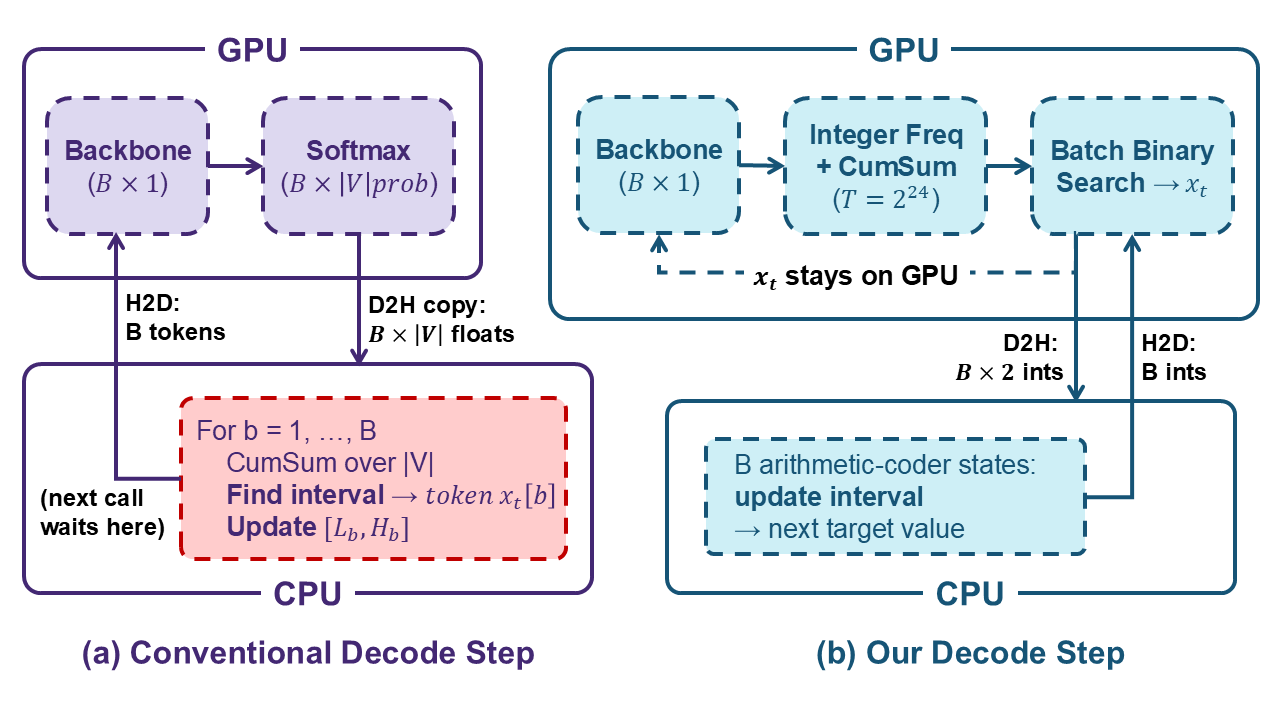}
\caption{One decoding step for batch size $=B$. (a)~The conventional pipeline copies the full probability matrix ($B \times |\mathcal{V}|$ floats) to the host, where each stream's arithmetic decoder builds a cumulative distribution over the vocabulary and recovers its token; the next backbone call waits for this CPU work (red block). (b)~Our pipeline quantizes probabilities and builds cumulative frequencies on the GPU. The host sends only each stream's target value ($B$ integers). With Rank-Tier Coding, (b) needs no device-to-host transfer.}
\label{fig:vs.pipeline}
\end{figure}

\paragraph{The conventional pipeline.}
Figure~\ref{fig:vs.pipeline} shows one decoding step of the pipeline we start from, the public implementation accompanying~\cite{waiting2026cidr}. The model produces logits on the GPU, which are turned into probabilities over the vocabulary ($|\mathcal{V}| \approx 151$K for Qwen2.5~\cite{qwen2025qwen25}). The probability vectors are copied to the host, where the arithmetic decoder builds each stream's cumulative distribution, locates the encoded value in it, and recovers the next token, which is the input to the next model call. With $N$ tokens per stream, decoding takes
\begin{equation}
T_{\text{dec}} \approx N \cdot \big(t_{\text{call}}(B) + t_{\text{host}}(B)\big),
\label{eq:tdec}
\end{equation}
where implementations can overlap part of $t_{\text{host}}$ with GPU execution~\cite{waiting2026cidr}, but not all of it, because the next call needs the tokens that the host has just decoded.
 
\paragraph{Two limits.}
Equation~\eqref{eq:tdec} contains two fundamentally different kinds of cost.
\emph{Execution redundancy} is work that the coded output does not require. The host receives $B \times |\mathcal{V}|$ probabilities per step although it needs only a few numbers per stream to recover each token, and it builds full cumulative distributions on the CPU; the encoder mirrors the decoder and calls the model once per token although all tokens are known. None of this affects which bits are produced. In the conventional pipeline it dominates: CPU-side arithmetic coding and the device-to-host copy of probability vectors together account for 68\% of decode time on Wikipedia text and 60\% on source code, more than the model itself (Section~\ref{subsec:time}).
\emph{Sequential dependency} is the factor $N$. It is a property of lossless decoding, not of an implementation: to recover $x_{t+1}$, the decoder needs a distribution conditioned on $x_{\le t}$, which it knows only after decoding $x_t$.
A third factor, the cost of each call, depends on model size, precision, and hardware. It is complementary to both limits.
 
\subsection{Addressing the Two Limits}
\label{subsec:price}
 
The two limits differ in what it costs to address them. Execution redundancy can be removed without changing the output, provided encoder and decoder still produce bit-identical predictions although removing redundancy makes them call the model with different shapes; Section~\ref{subsec:parity} shows why this does not hold by default and how we obtain it. Sequential dependency cannot be removed by better execution, only traded, for more calls or for more bits.
 
\paragraph{Fewer sequential calls cost bits under per-offset prediction.}
Consider multi-token coders of the form we study~\cite{stern2018blockwise, gloeckle2024multitoken, cai2024medusa}: a single model call resolves tokens $x_{t+1}, \dots, x_{t+n}$ by fixing, before any of them is decoded, a separate distribution $q_k(\cdot \mid x_{\le t})$ for each offset $k$, conditioned only on the prefix before the round. For any such coder, the expected code length of a round is at least
\begin{equation}
\mathbb{E}[\ell_{\text{round}}] \;\ge\; \sum_{k=1}^{n} H\big(X_{t+k} \mid X_{\le t}\big),
\end{equation}
whereas token-by-token decoding conditions each token on everything before it and can reach, by the chain rule, $\sum_{k} H(X_{t+k} \mid X_{< t+k}) = H(X_{t+1:t+n} \mid X_{\le t})$. The difference is
\begin{equation}
\Delta_n = \sum_{k=2}^{n} I\big(X_{t+k};\, X_{t+1:t+k-1} \,\big|\, X_{\le t}\big) \;\ge\; 0,
\label{eq:delta}
\end{equation}
the information that the earlier tokens of a round carry about the later ones, which per-offset prediction leaves unused. For this class of coders, $\Delta_n$ is a property of the coding scheme rather than of how well the predictors are trained: no choice of $q_k$ removes it, and a practical predictor additionally pays the divergence between $q_k$ and the true conditionals. The bound does not cover designs in which the distribution for a later offset depends on tokens decoded earlier in the same round, for example through verification of drafted tokens or structured joint prediction~\cite{leviathan2023speculative, chen2023speculative, ankner2024hydra}. Such designs reintroduce dependencies within the round or require modeling joint distributions over several tokens; we do not study them here.
 
\paragraph{The alternative is to pay in calls.}
A decoder can instead keep full conditioning, so that the decoding schedule imposes no conditioning loss, and make some sequential calls cheaper, for example by letting a smaller model handle the tokens it predicts correctly~\cite{leviathan2023speculative}; every token still requires one sequential call. After removing execution redundancy (Section~\ref{sec:substrate}), we study one technique at each end of this trade-off: SDMC pays in calls (Section~\ref{sec:speculative}), and MSMC pays in bits (Section~\ref{sec:msmc}).
\section{Removing Execution Redundancy}
\label{sec:substrate}
\label{sec:rank_compression}

This section removes the execution redundancy of Section~\ref{subsec:time-goes} without changing what is coded. Adams et al.~\cite{adams2026highthroughput} substantially reduced compression-side overhead through target-interval capture and pipelined arithmetic coding. Two questions remain open: whether block-parallel encoding stays lossless under half-precision arithmetic, and how to remove the redundancy in decompression, which their work does not evaluate. We first obtain bit-identical predictions across call shapes (Section~\ref{subsec:parity}), then build a pipeline that encodes whole blocks and keeps vocabulary-sized work on the GPU in both directions, for arithmetic coding and for Rank-Tier Coding (Section~\ref{subsec:rank_tier}).

\subsection{Numerical Parity Across Call Shapes}
\label{subsec:parity}
The encoder knows the complete input, so it can evaluate many consecutive positions in one teacher-forced forward pass~\cite{williams1989teacher}, whereas the decoder must evaluate one position per call. Lossless decoding requires both to produce the same prediction at every position: identical Top-$K$ lists for a rank-based coder and, more strictly, identical integer cumulative frequencies for arithmetic coding. A single mismatch, $p_t^{\mathrm{enc}} \ne p_t^{\mathrm{dec}} \Rightarrow \hat{x}_t \ne x_t$, corrupts every later token, since $\hat{x}_t$ is the input to all later positions.

Such mismatches arise because floating-point addition is not associative~\cite{goldberg1991floating}: $(a + b) + c \;\ne\; a + (b + c)$. In FP16, for example, $(2048 + 1) - 2048 = 0$, because $2049$ is not representable and rounds to $2048$, whereas $2048 + (1 - 2048) = 1$. Every matrix multiplication, attention score, and normalization in a transformer is a long sum, and standard GPU kernels choose how to split and order that sum, through tiling, split reductions, and the choice of tensor-core path, based on the shape of the call~\cite{he2025nondeterminism}. The same position can therefore produce slightly different logits when it is evaluated inside a teacher-forced block than when it is evaluated alone. These differences affect only the last bits, but they are enough to swap two near-tied candidates, and over millions of tokens near-ties occur. Fixing the batch size, as in~\cite{waiting2026cidr}, makes one execution schedule repeatable; it does not make an encoder that evaluates blocks agree with a decoder that evaluates single positions. Block-parallel encoding is therefore safe only if every reduction is performed in the same order regardless of the call shape.

\paragraph{Batch-invariant kernels.}
Following the design principle of He et al.~\cite{he2025nondeterminism}, we replace every operation whose reduction order could depend on the call shape with a Triton~\cite{tillet2019triton} kernel in which the computation for one position is independent of the others in the call:

\begin{itemize}[leftmargin=1.2em, itemsep=2pt, parsep=0pt]
\item[-] \textbf{Attention.} Each program computes one query row of one head in FP32 with an online softmax~\cite{milakov2018online, dao2022flashattention}, visiting fixed-size key blocks in a fixed order from the first cached position. Key blocks beyond a query's causal horizon, which a teacher-forced block adds, are fully masked and leave the accumulator bit-for-bit unchanged.
\item[-] \textbf{Linear projections.} All linear layers, including the LM head and the prediction heads of Section~\ref{sec:msmc}, use a fixed-tile matrix multiplication: rows are padded to a fixed tile height, and each tile is accumulated in FP32 in a fixed order on FP16 tensor cores.
\item[-] \textbf{RMSNorm~\cite{zhang2019rmsnorm}.} Each program normalizes one row in FP32, so the reduction over the hidden dimension is the same for every call shape.
\end{itemize}

We also disable TF32 and reduced-precision reductions in the remaining PyTorch matrix operations. Encoder and decoder run the same kernels, and we verify both at the kernel level, by comparing block evaluation against position-by-position evaluation bit for bit, and end to end (Section~\ref{subsec:setup}).

\paragraph{Shared schedule.}
The remaining sources of divergence are controlled by construction. The encoder replays the decoder's KV-cache schedule exactly: whenever the decoder's cache reaches the context length (1024 tokens), both sides truncate it to the most recent 512 entries at the same step, and the encoder never lets a block span such a truncation point. Every position is therefore evaluated on the same cached prefix on both sides. Probability quantization (for AC) and Top-$K$ selection (for Rank-Tier Coding) are applied to tensors of the same shape on both sides, so bit-identical logits yield the same frequencies and the same candidate ordering, including the order of tied candidates.

\paragraph{Arithmetic Coding.} Both sides convert logits to integer frequencies on the GPU with a fixed total $T{=}2^{24}$:
\begin{equation}
f_v = 1 + \big\lfloor p_v\,(T - |\mathcal{V}|) \big\rfloor, \qquad v \in \mathcal{V},
\end{equation}
assigning the rounding remainder to the most probable token so that the frequencies sum exactly to $T$, and take a cumulative sum over the vocabulary. Every token thus receives a nonzero frequency, and the quantization loses little probability mass. The coder itself is a standard 32-bit integer arithmetic coder~\cite{witten1987arithmetic} that runs on the CPU, one instance per stream.

\paragraph{Encoding.} For each position in a block, the GPU gathers the target token's interval $[\mathrm{cum}_{x_t} - f_{x_t}, \mathrm{cum}_{x_t})$ for all streams. Only these two integers per token are copied to the host, once per block, and a background thread feeds them to the per-stream encoders while the GPU processes the next block. This corresponds to the target-interval capture and pipelined coding of~\cite{adams2026highthroughput}.
\paragraph{Decoding.} At each step, every stream's decoder computes its current target value from its coder state, a single integer. The $B$ values are sent to the GPU, which locates each one in the corresponding cumulative frequencies with a batched binary search, recovers the token, and returns only the token's interval ($B \times 2$ integers) so that the host can update its coder states. The decoded tokens remain on the GPU as the input to the next call. This applies the target-interval principle to decoding, where the interval is not known in advance and must be located from the coder state.

Compared with the conventional pipeline, which copies a $B \times |\mathcal{V}|$ probability matrix to the host and builds each stream's cumulative distribution on the CPU at every step, the host now handles $O(B)$ integers per step instead of $O(B|\mathcal{V}|)$ floating-point values.

\paragraph{Rank-Tier Coding.} The same schedule applies, with rank computation in place of frequency quantization (Section~\ref{subsec:rank_tier}). The encoder computes Top-$K$ lists and ranks for a whole block on the GPU and passes ranks (and, for escapes, token IDs) to a background serializer. The decoder needs no device-to-host transfer at all: it parses the stored tier codes, sends them to the GPU, and recovers all $B$ tokens with one gather from the Top-$K$ lists.

Section~\ref{subsec:time} measures the effect of these changes: with the same model and probabilities, host-side work falls from the dominant cost of decoding to under 10\% of decode time. What remains is the model call itself, once per token: sequential dependency.

\subsection{Rank-Tier Coding}
\label{subsec:rank_tier}

Arithmetic coding assigns each token a code length close to $-\log_2 P(x_t \mid x_{<t})$, which requires building a cumulative distribution and updating an interval at every step. Rank-Tier Coding instead encodes each token by its position in the model's ranked predictions, a form of symbol ranking that dates back to Shannon~\cite{shannon1951prediction} and has been paired with LLM predictors before, with the rank stream passed to a general-purpose compressor~\cite{valmeekam2023llmzip, finezip} or coded with bit-packing or Huffman codes~\cite{adams2026highthroughput}. We use a simple fixed-length tier code computed on the GPU, which requires only a Top-$K$ selection and a comparison per token. We include it as the cheapest coder we could pair with the pipeline, to measure how much the choice of coder still matters once redundancy is removed.

\paragraph{Code.} Let $z_t = \mathcal{M}(x_{<t}) \in \mathbb{R}^{|\mathcal{V}|}$ be the logits for position $t$, and $I_t = (I_t[1], \ldots, I_t[K])$ the Top-$K$ ($K=32$) token IDs sorted by decreasing logit with deterministic tie-breaking. The rank $r_t$ of $x_t$ is its 1-indexed position in $I_t$; if $x_t \notin I_t$, then $r_t > K$ and the token is encoded as an escape, analogous to the escape mechanism of PPM~\cite{cleary1984ppm} and the exceptions of thresholded symbol ranking~\cite{rankthreshold2025}. Ranks map to four tiers:
\begin{equation}
T(r_t) =
\begin{cases}
1, & r_t = 1,\\
2, & 2 \le r_t \le 4,\\
3, & 5 \le r_t \le 32,\\
4, & r_t > 32.
\end{cases}
\end{equation}
For the first three tiers, the payload is $r_t-1$, which fits in 2 bits for Tier~2 and 5 bits for Tier~3. Tier~4 stores the raw token ID in 18 bits, since $|\mathcal{V}| \approx 1.5\times 10^{5} \le 2^{18}$:
\begin{equation}
\mathcal{P}(r_t,x_t) =
\begin{cases}
\emptyset, & T(r_t)=1,\\
\mathrm{Bin}(r_t-1,2), & T(r_t)=2,\\
\mathrm{Bin}(r_t-1,5), & T(r_t)=3,\\
\mathrm{Bin}(x_t,18), & T(r_t)=4.
\end{cases}
\end{equation}
With a 2-bit tier label, the code length of a token is $\ell_t = 2 + |\mathcal{P}(r_t,x_t)|$, i.e., 2, 4, 7, or 20 bits for Tiers 1--4.

\paragraph{Decoding.} The decoder reconstructs the same Top-$K$ list $I_t$ from the decoded prefix and recovers the token as
\begin{equation}
\hat{x}_t =
\begin{cases}
I_t[\rho(T,\mathrm{Payload})], & T \in \{1,2,3\},\\
\mathrm{Payload}, & T=4,
\end{cases}
\end{equation}

\begin{equation}
\rho(T,\mathrm{Payload}) =
\begin{cases}
1, & T=1,\\
\mathrm{Payload}+1, & T\in\{2,3\}.
\end{cases}
\end{equation}
Decoding is lossless whenever encoder and decoder produce the same Top-$K$ list at every position, which Section~\ref{subsec:parity} guarantees. The argument is inductive: if the decoded prefix matches the original, both sides compute the same list and recover the same next token.

\paragraph{What the fixed code gives up.} Rank-Tier Coding needs neither a probability vector nor a coder state, so its per-token work is a Top-$K$ selection and a comparison, and its encoding of each position is independent of the others. The price is that code lengths are fixed by tier rather than by the model's confidence: a correct top-1 prediction always costs 2 bits, whereas arithmetic coding can spend a small fraction of a bit on a confident prediction. Rank-Tier Coding therefore cannot reach the bound of Equation~\eqref{eq:codelength}, and its compression rate is bounded below by 2 bits per token regardless of model quality. Section~\ref{sec:experiments} quantifies this trade-off: on the same pipeline, Rank-Tier Coding decodes 10--20\% faster than arithmetic coding but costs 0.4--0.8 more bits per byte. Once redundancy is removed, a cheaper coder therefore buys little speed at a large price in compression.
\section{Speculative Dual-Model Coding}
\label{sec:speculative}

Once execution redundancy is removed (Section~\ref{sec:substrate}), decoding is dominated by one sequential model call per token. Section~\ref{subsec:price} showed that, with a fixed model, this dependency can be paid for in calls or in bits. This section studies the first option: keep every token’s full conditioning, so that the decoding schedule itself costs no bits, and make some of the sequential calls cheaper. How many bits SDMC spends then depends on how it codes each token. Speculative decoding~\cite{leviathan2023speculative, chen2023speculative, xia2024speculative} does this in text generation by letting a small draft model propose tokens for a larger target model; Adams et al.~\cite{adams2026highthroughput} name it as a promising direction for compression. We study whether it helps. Compression has an advantage over generation here: the encoder knows the true sequence and can record whether each draft prediction is correct, so the decoder never needs to verify a draft. We call this design Speculative Dual-Model Coding (SDMC).

\subsection{Design}
\label{SDMC_design}

SDMC pairs a small draft model $M_{\text{S}}$ (Qwen2.5-0.5B) with a larger target model $M_{\text{L}}$ (Qwen2.5-7B)~\cite{qwen2025qwen25}; both share a tokenizer. For each token $x_{i+1}$, the encoder first checks whether it is the draft model's top-1 prediction. If so, it emits a single flag bit. Otherwise, it emits a zero flag followed by the token's Rank-Tier code under the target model (Section~\ref{subsec:rank_tier}):
\begin{equation}
\text{encode}(x_{i+1}) =
\begin{cases}
\texttt{1} & \text{if } \text{rank}_{M_{\text{S}}}(x_{i+1}) = 1,\\
\texttt{0}\,\|\,T(r)\,\|\,\mathcal{P}(r, x_{i+1}) & \text{otherwise, with } r = \text{rank}_{M_{\text{L}}}(x_{i+1}).
\end{cases}
\end{equation}
A hit thus costs 1 bit and a miss at least 3 bits. With draft hit rate $p_{\text{S}}$, the expected cost per token is
\begin{equation}
\bar{b} = p_{\text{S}} + (1 - p_{\text{S}}) \Big[ 3 + \sum_{k=2}^{4} \hat{p}_k^{\text{L}}\, \delta_k \Big],
\end{equation}
where $\hat{p}_k^{\text{L}}$ is the probability that a missed token falls into tier $k$ of $M_{\text{L}}$ and $\delta_k \in \{2, 5, 18\}$ is that tier's payload.

\paragraph{Encoding.} The encoder knows all tokens, so $M_{\text{L}}$ evaluates blocks of $\gamma$ positions per call, as in Section~\ref{subsec:parity}. $M_{\text{S}}$ is stepped one position at a time to match the decoder, which uses it to generate tokens. Both models run on every token, for $B$ streams in parallel.

\paragraph{Decoding.} The decoder splits each stream into maximal \emph{runs}: an S-run of consecutive hits or an L-run of consecutive misses. In an S-run of length $L$, $M_{\text{S}}$ generates the $L$ tokens one call at a time from its own top-1 predictions, and $M_{\text{L}}$ then ingests them in one prefill call to update its KV cache~\cite{pope2023efficiently}. In an L-run, $M_{\text{L}}$ decodes the tokens one call at a time from their stored ranks, and $M_{\text{S}}$ then catches up in one prefill call. Each token is processed by both models, but only one of them needs a sequential single-token call for it. The two KV caches therefore advance on offsetting schedules that depend on the data, and these schedules differ across streams. In our implementation, streams are therefore decoded one at a time ($B{=}1$).

\paragraph{Adapting the draft model.} The draft hit rate $p_{\text{S}}$ determines both the code length and how often $M_{\text{L}}$ is skipped; each additional percentage point of $p_{\text{S}}$ saves at least 0.02 bits per token, because a miss costs at least 3 bits against 1 for a hit. We therefore also tested fine-tuning $M_{\text{S}}$ with LoRA~\cite{hu2022lora}, using cross-entropy plus a hinge term~\cite{crammer2001multiclass} that penalizes the target token when its logit does not exceed that of its strongest competitor by at least a margin $\Delta$:
\begin{equation}
\mathcal{L} = -\frac{\alpha}{N} \sum_{i=1}^{N} \log q_i \;+\; \frac{\beta}{N} \sum_{i=1}^{N} \max\!\Big(0,\; \Delta + \max_{j \neq y_i} z_j^{(i)} - z_{y_i}^{(i)}\Big),
\end{equation}
where $q_i$ is the draft model’s probability of the target token $y_i$.
Fine-tuned on the target file, this raises the draft's top-1 hit rate on Wiki from 47.7\% to 55.3\% with cross-entropy alone and to 56.5\% with the hinge term. Because these adapters are specific to a file, their size would have to be counted toward the compressed output; the comparison in Section~\ref{subsec:sdmc-vs-msmc} therefore uses the un-adapted base model.

\subsection{Limitations of SDMC}
\label{subsec:sdmc_limits}

\paragraph{Savings are capped by the hit rate and the run structure.}
SDMC does not reduce computation: every token passes through both models. It only replaces some sequential single-token calls to $M_{\text{L}}$ with calls to $M_{\text{S}}$. Let $t_{\text{S}}$ and $t_{\text{L}}$ be the latencies of a single-token call, $t^{\text{pf}}_{\text{S}}$ and $t^{\text{pf}}_{\text{L}}$ those of a prefill call, $R_{\text{S}}$ and $R_{\text{L}}$ the numbers of S- and L-runs, and $N_{\text{S}} = p_{\text{S}} N$, $N_{\text{L}} = (1-p_{\text{S}}) N$. By analogy with the analysis of speculative decoding~\cite{leviathan2023speculative}, decoding takes approximately
\begin{equation}
\begin{gathered}
T_{\text{SDMC}} \approx N_{\text{S}}\, t_{\text{S}} + N_{\text{L}}\, t_{\text{L}} + R_{\text{S}}\, t^{\text{pf}}_{\text{L}} + R_{\text{L}}\, t^{\text{pf}}_{\text{S}} \\[1ex]
\text{versus}\qquad
T_{\text{base}} = N\, t_{\text{L}}.
\end{gathered}
\end{equation}
A large speedup requires a high hit rate, long runs, and $t_{\text{S}} \ll t_{\text{L}}$ at the same time. None of these holds comfortably. The hit rate is bounded by how predictable the data is; hits and misses interleave, so runs are short and each switch adds a prefill call; and at $B{=}1$, a single-token call is dominated by fixed per-call latency rather than arithmetic~\cite{pope2023efficiently}, so a call to the 0.5B model is far less than $14\times$ cheaper than a call to the 7B model.

\paragraph{SDMC adds a sequential loop to encoding.}
On the encoding side, the target model already processes blocks, so there are no sequential calls to save. The draft model, however, must be stepped one position at a time to reproduce the decoder's computation. SDMC therefore adds a sequential loop to an encoder that would otherwise have none. Whether this makes encoding slower in absolute terms depends on how fast the single-model coder encodes (Section~\ref{subsec:sdmc-vs-msmc}).

\paragraph{Half-precision parity is not reached in our implementation.}
Lossless decoding requires the decoder to reproduce the draft model's top-1 decisions and the target model's candidate lists exactly. In single-model coding, encoder and decoder differ in a fixed, known way, the number of positions per call, which the kernels of Section~\ref{subsec:parity} handle. In SDMC, each model's KV cache is written by single-token calls in some runs and by prefill calls of data-dependent length in others, and the encoder writes it differently again. With FP16 or BF16 execution, our implementation fails to reconstruct some inputs, and it still fails when we apply the batch-invariant kernels to both models. The failure persists in a control in which $M_{\text{S}}$ and $M_{\text{L}}$ are the same model, which leaves only the coordination pattern as a difference. We hypothesize that caches written under data-dependent call shapes are the source, but we have not located the first diverging operation, so we report this as a limitation of our implementation rather than of dual-model coordination in general. We therefore run both models in FP32 for all SDMC experiments, which makes decoding lossless on every input we test but forgoes half-precision tensor cores~\cite{micikevicius2018mixed}.

\paragraph{Lesson.}
SDMC saves sequential calls only in proportion to the draft's hit rate, slows encoding, and in our implementation requires FP32 and single-stream decoding. These costs share a cause: progress per sequential call depends on a second model whose state must be kept in step with the first. In the terms of Section~\ref{subsec:price}, SDMC pays for sequential dependency in calls, and since a call to a small model is not much cheaper than one to a large model when decoding is latency-bound, the payment buys little. Section~\ref{sec:msmc} studies the other option, fewer calls paid for in bits. Section~\ref{subsec:sdmc-vs-msmc} reports the measurements.
\section{Multi-Token Single-Model Coding}
\label{sec:msmc}

SDMC pays for sequential dependency in calls. Multi-Token Single-Model Coding (MSMC) takes the other option of Section~\ref{subsec:price}: it makes fewer calls and pays in bits, by letting one model resolve several tokens per call. Following Medusa~\cite{cai2024medusa} and related multi-token predictors~\cite{stern2018blockwise, gloeckle2024multitoken}, it attaches lightweight heads to a frozen backbone, each predicting a different future offset from the same hidden state. Encoding and decoding both run on the pipeline of Section~\ref{sec:substrate}, and MSMC works with either coder: each predicted distribution is coded with arithmetic coding or Rank-Tier Coding exactly as a single-token prediction would be. The number of tokens resolved per call, $n$, is chosen at encoding time. We describe the design (Section~\ref{subsec:msmc_framework}), quantify what $n$ costs in compression (Section~\ref{subsec:msmc_bits}) and gains in speed (Section~\ref{subsec:msmc_complexity}), and combine both into a model for choosing $n$ (Section~\ref{subsec:msmc_choose}). MSMC does not compress better than token-by-token arithmetic coding. It adds operating points with higher decode throughput at a known cost in bits, which are useful when decoding latency matters more than compression rate.

\subsection{Design}
\label{subsec:msmc_framework}

\paragraph{Heads.}
Let $\mathbf{h}_t \in \mathbb{R}^{d}$ be the backbone's final (post-norm) hidden state at position $t$. Offset $+1$ is predicted by the frozen backbone LM head $W^{lm}_{\text{base}}$, so MSMC with $n{=}1$ is exactly single-model coding. For offsets $+2$ to $+K{+}1$ we train $K=5$ heads: $H_1, \dots, H_K$, each a feed-forward block with a residual connection~\cite{he2016resnet, elfwing2018silu}:
\begin{equation}
\begin{gathered}
  H_k(\mathbf{h}) = W_k^{lm}\Big(\mathrm{SiLU}\big(W_k^{ff}\,\mathbf{h}\big) + \mathbf{h}\Big), \\[1ex]
  \mathrm{softmax}\big(H_k(\mathbf{h}_t)\big) \approx P(x_{t+k+1} \mid x_{\le t})
\end{gathered}
\end{equation}
with $W_k^{ff} \in \mathbb{R}^{d \times d}$ and $W_k^{lm} \in \mathbb{R}^{|\mathcal{V}| \times d}$. Following Medusa, $W_k^{ff}$ is initialized to zero, and we initialize $W_k^{lm}$ as a copy of $W^{lm}_{\text{base}}$, so every head starts as the backbone's next-token predictor and then adapts to its offset.

\begin{figure}[t]
  \centering
  \includegraphics[width=\linewidth]{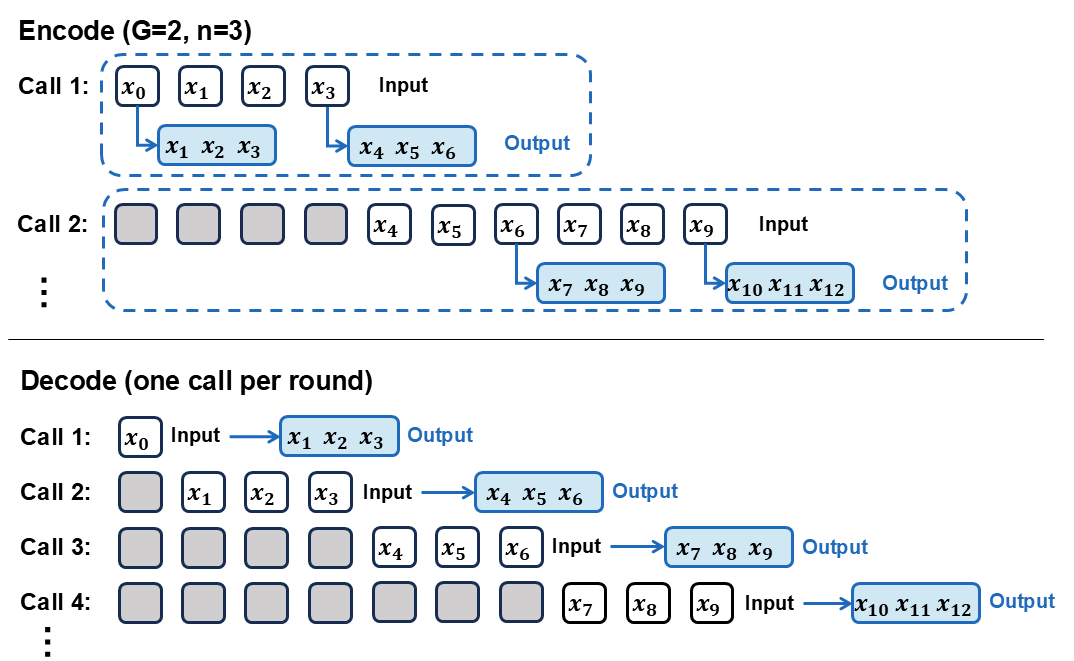}
  \caption{MSMC schedule for one stream (let $n{=}3$, $G{=}2$). From each predictor position $t \in \{0, n, 2n, \dots\}$, the backbone LM head predicts $x_{t+1}$ and heads $H_1, \dots, H_{n-1}$ predict $x_{t+2}, \dots, x_{t+n}$. Encoding covers $G$ rounds per call by teacher forcing, so the last $n$ targets of one call are fed in the next; decoding covers one round per call. Gray boxes are positions already in the KV cache. Once the cache reaches the context length, both sides truncate it at the same round (not shown). All $B$ streams follow this schedule in lockstep.}
  \label{fig:MSMC-process}
\end{figure}

\paragraph{Encoding.}
A \emph{round} resolves $n$ consecutive tokens from one predictor position $t \in \{0, n, 2n, \dots\}$: from $\mathbf{h}_t$, the backbone LM head predicts $x_{t+1}$ and heads $H_1, \dots, H_{n-1}$ predict $x_{t+2}, \dots, x_{t+n}$. Because the input is known, the encoder covers $G$ rounds, that is, $Gn$ coded tokens, in one teacher-forced backbone call. Each call feeds the tokens up to the last predictor position of its rounds, so the final $n$ targets of one call are fed as input in the next (Figure~\ref{fig:MSMC-process}). Attention reads all earlier positions from the KV cache, subject to the truncation schedule of Section~\ref{subsec:parity}, and the causal mask ensures that $\mathbf{h}_t$ depends only on $x_{\le t}$, exactly as in decoding. The LM head and the heads are evaluated only at predictor positions, and each prediction is coded with arithmetic coding or Rank-Tier Coding as in Section~\ref{subsec:parity}.

\paragraph{Decoding.}
\label{subsec:msmc-decode}
The decoder advances all $B$ streams one round per backbone call. The call for predictor position $t$ ingests the $n$ tokens $x_{t-n+1}, \dots, x_t$ recovered in the previous round (only $x_0$ for $t{=}0$; the first token of each stream is stored in the header), appends them to the KV cache, and returns $\mathbf{h}_t$; the LM head and the heads then produce the $n$ distributions from which the coder recovers $x_{t+1}, \dots, x_{t+n}$. Because these tokens are unknown until they are decoded, the decoder cannot cover several rounds in one call, so each round costs exactly one sequential backbone call. The value of $n$ is stored in the stream header.

\paragraph{Training per domain.}
Heads are trained once per data domain, with the backbone frozen, by minimizing the cross-entropy of each head at its offset, averaged over heads:
\begin{equation}
\begin{gathered}
\mathcal{L} = \frac{1}{K}\sum_{k=1}^{K} \mathcal{L}_k, \\[1ex]
\mathcal{L}_k = -\frac{1}{N-k-1} \sum_{t=0}^{N-k-2} \log \hat{P}_k(x_{t+k+1} \mid x_{\leq t}).
\end{gathered}
\end{equation}
The training data come from the same domain as the data to be compressed but exclude it; in our evaluation, they are held-out files of each corpus, disjoint from the test files (Section~\ref{subsec:setup}). Training on the input itself would make the heads part of the compressed representation, since under the minimum description length principle~\cite{rissanen1978modeling} the cost of a model fitted to the data counts toward its code length; training per domain makes them share infrastructure like the backbone.

\paragraph{Storage and its amortization.}
Each head has $d^2 + d|\mathcal{V}|$ parameters, dominated by its vocabulary projection: for the 0.5B backbone ($d{=}896$, $|\mathcal{V}| \approx 1.5\times10^5$), about $1.4\times10^8$ parameters, or 0.27\,GB in FP16. Five heads thus occupy $1.37$\,GB, which is comparable to or larger than the backbone itself. Because one set of heads serves a whole domain, its storage $S_{\text{heads}}$ can be amortized over the volume $D$ of data compressed with it, adding
\begin{equation}
\Delta\text{bpb} = \frac{8\, S_{\text{heads}}}{D}
\end{equation}
bits per byte. For heads of 1.4\,GB, this is 0.011\,bpb at $D = 1$\,TB and 0.11\,bpb at $D = 100$\,GB. MSMC therefore suits domains with large volumes of data, such as the archival settings of Section~\ref{sec:intro}, and does not suit small, one-off inputs. Sharing $W^{lm}_{\text{base}}$ across heads would reduce the storage to $d^2$ per head; we leave this to future work.

\paragraph{Losslessness.}
Encoder and decoder evaluate the same predictor positions on the same cached prefixes, but with different numbers of positions per call ($Gn$ versus $n$). The batch-invariant kernels of Section~\ref{subsec:parity}, which also cover the LM head and the heads, make the resulting distributions bit-identical, so decoding is lossless with either coder. Losslessness does not depend on how well the heads predict: a poor prediction only produces a longer code (a lower-probability interval, or a deeper tier or escape), never a wrong token.

\subsection{What Fewer Calls Cost in Compression}
\label{subsec:msmc_bits}

In a round of $n$ tokens, the token at offset $+k$ is coded with a distribution conditioned only on the prefix before the round, so by Section~\ref{subsec:price}, MSMC pays at least $\Delta_n$ extra bits per round compared with token-by-token decoding under the true distribution, plus the gap between its heads' predictions and the true conditionals. Both parts are captured by a quantity we can measure directly: the average code length $\ell_k$ that the coder spends on tokens at offset $+k$. Since each round codes exactly one token at each offset $1, \dots, n$, the compression rate at round size $n$ is
\begin{equation}
\text{bpb}(n) \;\approx\; \frac{1}{\beta}\cdot\frac{1}{n}\sum_{k=1}^{n} \ell_k,
\label{eq:bpb-n}
\end{equation}
where $\beta$ is the average number of bytes per token. Equation~\eqref{eq:bpb-n} separates what is fixed from what can be improved. $\ell_1$ is the code length of the backbone itself and does not depend on $n$. Under the true distribution, each further offset would add at least the information lost by not conditioning on the $k-1$ preceding tokens of the round. Our backbone and heads, however, both make prediction errors, so the measured gap $\ell_k - \ell_1$ mixes this conditioning loss, which no head can recover, with the prediction error of the head at offset $k$ relative to the backbone, which better-trained heads could reduce. Our measurements do not separate the two, so the measured loss should not be read as entirely unavoidable. Section~\ref{subsec:why-degrade} measures $\ell_k$ and shows that Equation~\eqref{eq:bpb-n} reproduces the measured bpb.

\subsection{What Fewer Calls Gain in Speed}
\label{subsec:msmc_complexity}


We compare MSMC with single-model coding on the same pipeline and coder, that is, with MSMC at $n{=}1$. One backbone call serves all $B$ streams, so we count calls per stream.

\paragraph{Sequential calls.}
Single-model decoding of $N$ tokens takes $N$ dependent calls, MSMC takes $N/n$:
\begin{equation}
T_{\text{base}} = N\, t_{\text{call}}(1),
\qquad
T_{\text{msmc}} = \frac{N}{n}\, t_{\text{call}}(n),
\qquad
S = n\,\frac{t_{\text{call}}(1)}{t_{\text{call}}(n)},
\end{equation}
where $t_{\text{call}}(n)$ is the latency of a call that ingests $n$ positions per stream and evaluates $n$ projections. The speedup approaches $n$ when $t_{\text{call}}(n) \approx t_{\text{call}}(1)$, that is, when a call is latency-bound: dominated by reading the weights and launching kernels rather than by arithmetic~\cite{pope2023efficiently, williams2009roofline}. This holds when $B$ is small. As $B$ grows, each call carries $Bn$ positions, becomes compute-bound, and $t_{\text{call}}(n)$ grows with $n$, so the speedup shrinks.

\paragraph{Total work.}
Per round, MSMC processes $n$ positions through the backbone body, one LM-head projection, and $n{-}1$ head evaluations, each costing $C_{\text{lm}} + d^2$. Per token, this is
\begin{equation}
C_{\text{body}} + C_{\text{lm}} + \frac{n-1}{n}\, d^2,
\end{equation}
essentially the same work as single-model coding, which evaluates the LM head at every position. MSMC thus reorganizes the same work into fewer, larger calls rather than reducing it. It helps only where the GPU would otherwise idle between dependent calls, and not in encoding, whose teacher-forced calls already keep the GPU busy: from $n{=}1$ to $n{=}6$, encode throughput changes by at most 13\% (Table~\ref{tab:main-results}), while decode throughput roughly doubles. At large $B$, two further costs reduce the gain: coder work scales with tokens rather than calls, and each call reads $n{-}1$ additional vocabulary projections. Section~\ref{subsec:when-pays} confirms these predictions.

\subsection{Choosing the Number of Tokens per Call}
\label{subsec:msmc_choose}

Equation~\eqref{eq:bpb-n} and the speedup $S(n, B) = n\,t_{\text{call}}(1)/t_{\text{call}}(n)$ describe the two sides of the trade-off with quantities that can be measured once per domain and hardware: per-offset code lengths $\ell_k$ from a sample of the data, and call latencies $t_{\text{call}}(n)$ at the intended batch size. Given a compression budget, for example to stay below the bpb of the best classical compressor, the operator picks the largest $n$ whose predicted bpb satisfies it, and the speedup that $n$ buys follows from the batch size. Two regimes follow. When many independent streams are available, batching alone fills the GPU, $S$ falls well below $n$, and a small $n$ captures most of the attainable speedup at a modest cost in bpb. When few streams are available, as in decompressing a single block or serving random access, batching cannot help, $S$ approaches $n$, and fewer calls are the only way to speed up decoding with a fixed model. Section~\ref{subsec:when-pays} quantifies both regimes.

\section{Experiments}
\label{sec:experiments}

We evaluate the framework of Sections~\ref{sec:substrate}--\ref{sec:msmc} along two axes: compression rate, measured in bits per byte (bpb), and encode/decode throughput, measured in KB/s of original data. The evaluation is organized around four questions:
(i)~how the resulting operating points compare with conventional LLM-based coding and with classical compressors (Section~\ref{subsec:overall});
(ii)~where decoding time is spent, and how much of the speedup comes from the execution pipeline rather than the choice of entropy coder (Section~\ref{subsec:time});
(iii)~why compression degrades as more tokens are resolved per backbone call (Section~\ref{subsec:why-degrade}); and
(iv)~in which execution regime multi-token decoding pays off (Section~\ref{subsec:when-pays}).
Section~\ref{subsec:sdmc-vs-msmc} then revisits the dual-model alternative of Section~\ref{sec:speculative}.

Overall, Section~\ref{subsec:overall} reports end-to-end results; Sections~\ref{subsec:time}--\ref{subsec:when-pays} analyze where the differences come from.

\subsection{Experimental Setup}
\label{subsec:setup}

\paragraph{Hardware.} All experiments run on a single NVIDIA H100 GPU (80\,GB HBM3) without multi-GPU parallelism or CPU offloading; all throughput numbers are single-device. Losslessness and throughput are measured on this GPU type.

\paragraph{Datasets.} We use two corpora with different statistical structure:

\begin{itemize}[leftmargin=1.2em, itemsep=2pt, parsep=0pt]
\item[-] \textbf{Wiki}: plain-text files from the \emph{Plain Text Wikipedia (Simple English)} dataset~\cite{kaggle_simplewiki}, which provides roughly 170 $\sim$1\,MB files. We randomly sample 10 files for evaluation.
\item[-] \textbf{Code}: $\sim$1\,MB excerpts of source code drawn from MBPP~\cite{austin2021mbpp}, HumanEval~\cite{chen2021humaneval}, and RocksDB~\cite{rocksdb}. We sample 10 excerpts for evaluation.
\end{itemize}

Every reported number is the mean over the 10 files of a corpus; per-file variation is shown as $\pm1$ standard deviation in Figure~\ref{fig:pareto}. To test longer streams, we also concatenate the ten Wiki files into one 10\,MB input (Table~\ref{tab:long-streams}).

\paragraph{Backbone and prediction heads.} All LLM-based methods use Qwen2.5-0.5B-Instruct~\cite{qwen2025qwen25} in FP16. Offset $+1$ is always predicted by the frozen backbone LM head; the MSMC heads predict offsets $+2$ to $+6$. Heads are trained once per \emph{domain}, not per file: one set for Wiki and one for Code, on files of that corpus that exclude the 10 evaluation files, which we verify by file hash and line-level overlap. Five of these files are held out for validation. The Wiki heads are trained on 160\,MB of text (about 22{,}000 steps) and the Code heads on 36\,MB (4{,}448 steps), each for 2 epochs on sequences of 1024 tokens with batch size 4, AdamW (learning rate $10^{-3}$ for the residual block and $10^{-4}$ for the vocabulary projection, cosine schedule with 5\% warmup), and BF16 mixed precision. This takes about 17 and 3.5 GPU-hours on a single H100, respectively. For each head, we keep the checkpoint with the lowest validation loss at its own offset. Each head has about $1.4\times10^8$ parameters (0.27\,GB in FP16; Section~\ref{subsec:msmc_framework}). This is a one-time cost per domain.

\paragraph{Accounting for model parameters.} We do not count the backbone or the domain heads toward bpb. Both are input-independent: the same weights serve every file of a domain, so they are deployed infrastructure in the same sense as the backbone, rather than part of any compressed artifact. Their one-time costs (training time and storage) are reported above and are amortized over all data of the domain. Every stream header needed for decoding \emph{is} counted.

\paragraph{Baselines.} \emph{Classical compressors}: gzip~(-9), zstd~(-3 and -19), and xz~(-9)~\cite{deutsch1996gzip, collet2021zstd, collin_xz}, run on the same files. \emph{Conventional AC.} The public implementation accompanying~\cite{waiting2026cidr}, in which each decoding step runs the backbone, copies the probability vector to the host, and updates the arithmetic coder on the CPU. Both pipelines thus use the same backbone and context schedule, and code with essentially the same probabilities.

\paragraph{Configurations.} Unless stated otherwise, all LLM-based methods use batch size $B{=}256$ for both encode and decode, where $B$ is the number of independent streams into which a file is split. We write MSMC\,+\,AC and MSMC\,+\,RC for multi-token decoding with Arithmetic Coding and Rank-Tier Coding, with $n$ tokens resolved per backbone call. $n{=}1$ uses no prediction heads and isolates the effect of the execution pipeline.

\paragraph{Metrics.} Throughput is reported in KB/s of original data rather than tokens/s, so that LLM-based and byte-oriented compressors share a unit. Throughput is the wall-clock time of the full encode or decode after the model is loaded. Reported bpb for all LLM-based methods, AC and Rank-Tier alike, includes the stream header.

\paragraph{Losslessness.} Every configuration in this section is verified to reconstruct the input byte-for-byte on all 20 evaluation files and on the 10MB concatenated input. Losslessness of the FP16 configurations relies on the batch-invariant kernels of Section~\ref{subsec:parity}; it holds when encoder and decoder run the same kernels, GPU architecture, and value of $n$. Encoding and decoding run as separate processes: the encoder writes the compressed file to disk, and the decoder, in a new process, reconstructs the input from that file and the saved backbone and head checkpoints alone, with no state carried over from encoding. Verification therefore exercises exactly the artifacts a deployment would use.

\subsection{How Do the Methods Compare Overall?}
\label{subsec:overall}

\begin{figure}[!t]
\centering
\begin{subfigure}[b]{0.49\columnwidth}
\centering
\includegraphics[width=\linewidth]{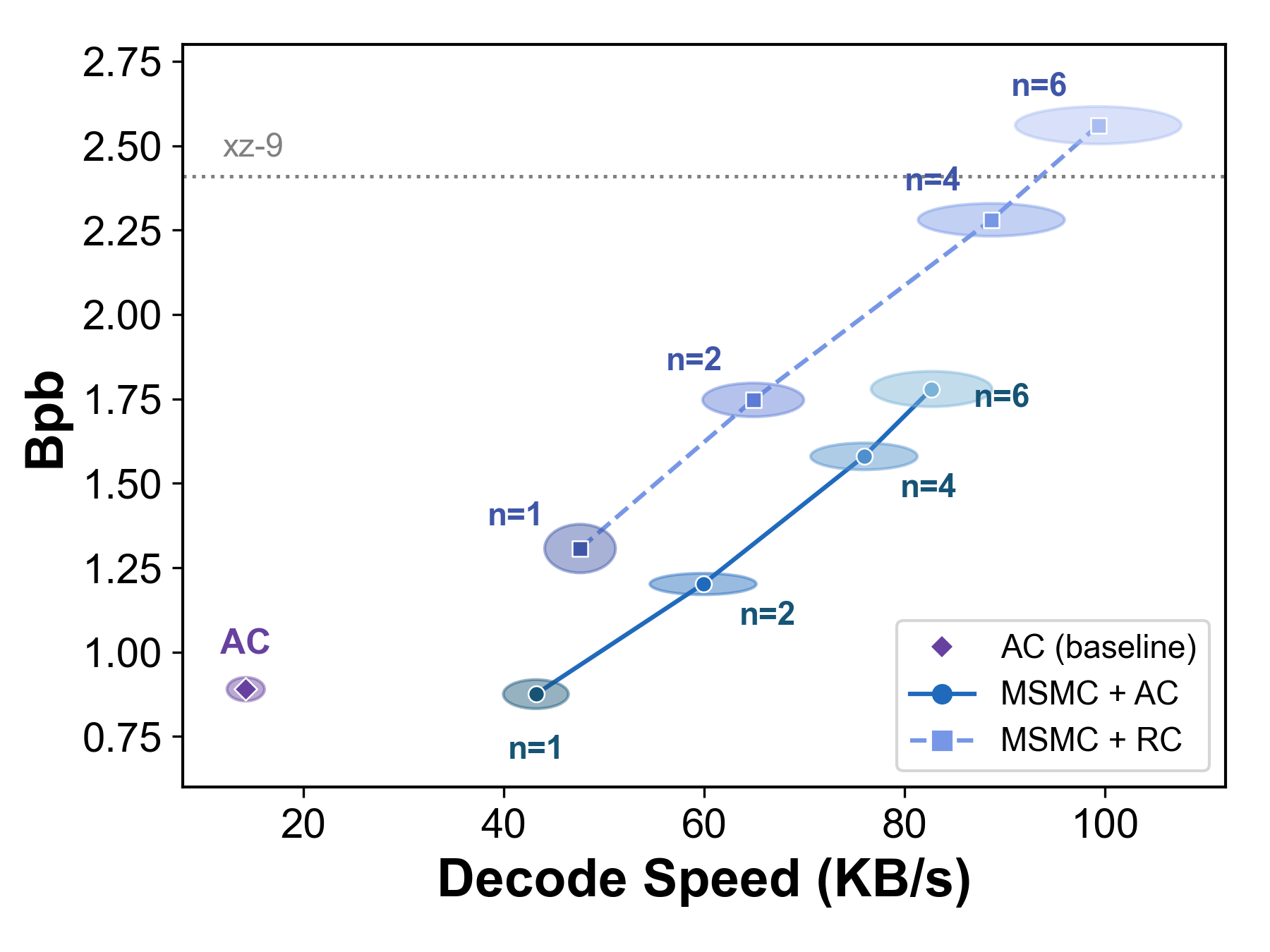}
\caption{Wiki}
\label{fig:pareto-wiki}
\end{subfigure}\hfill
\begin{subfigure}[b]{0.49\columnwidth}
\centering
\includegraphics[width=\linewidth]{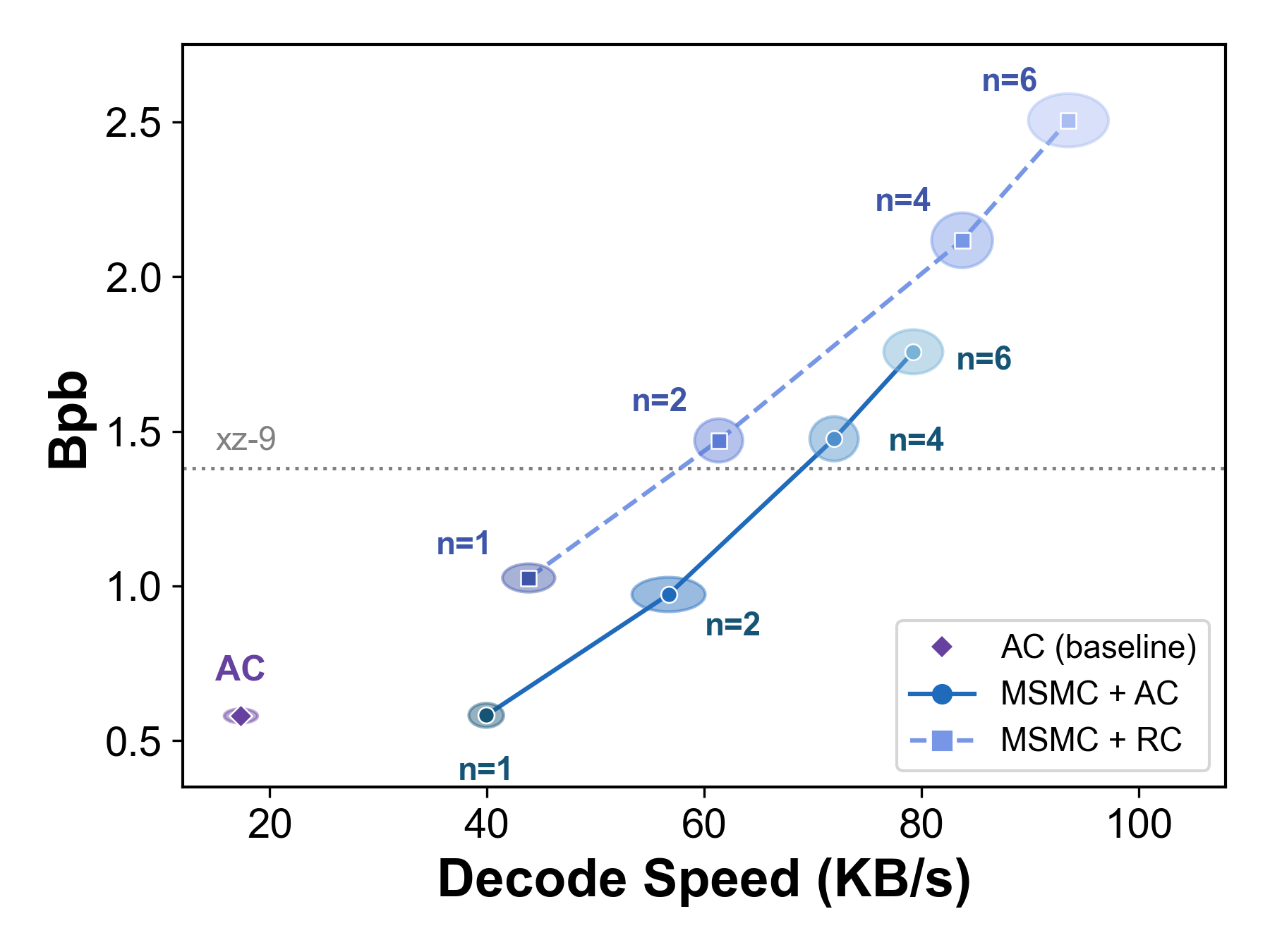}
\caption{Code}
\label{fig:pareto-code}
\end{subfigure}
\caption{Compression rate versus decode throughput for conventional AC, MSMC + AC, and MSMC + RC ($n{=}1,2,4,6$), same setting as Table~\ref{tab:main-results}, $n=3$ and $n=5$ omitted for space; trends are monotone. Ellipses span $\pm1$ standard deviation across the 10 files of each dataset. The dotted line marks \mbox{xz-9}.}
\label{fig:pareto}
\end{figure}

\begin{table*}[t]
\centering
\caption{Main results on Wiki and Code (mean over 10 files per dataset). LLM-based methods use Qwen2.5-0.5B-Instruct~\cite{qwen2025qwen25}, FP16, batch size $B{=}256$ for both encode and decode.
For MSMC, $n$ is the number of tokens resolved per backbone call; $n{=}1$ uses only the frozen backbone LM head.
MSMC rows run on our batch-invariant, teacher-forced pipeline.
Bpb of all LLM-based methods includes the stream header. Classical compressors are run on the same files.
\textit{Italic} bpb: worse than xz-9. All configurations are verified lossless at the byte level.}
\label{tab:main-results}
\small
\begin{tabular*}{\textwidth}{@{\extracolsep{\fill}} l rrr rrr}
\toprule
& \multicolumn{3}{c}{Wiki} & \multicolumn{3}{c}{Code} \\
\cmidrule(lr){2-4}\cmidrule(lr){5-7}
Method & Encode Speed & Decode Speed & Bpb ($\downarrow$) & Encode Speed & Decode Speed & Bpb ($\downarrow$) \\
\midrule
\multicolumn{7}{l}{\textit{Classical compressors}} \\
gzip-9   & 14.1 MB/s  & 199.3 MB/s  & 2.98 & 16.2 MB/s  & 302.5 MB/s  & 1.72 \\
zstd-3   & 207.1 MB/s & 1185.3 MB/s & 2.90 & 312.4 MB/s & 1430.1 MB/s & 1.75 \\
zstd-19  & 4.5 MB/s   & 1246.0 MB/s & 2.46 & 4.6 MB/s   & 1539.8 MB/s & 1.40 \\
xz-9     & 3.2 MB/s   & 106.7 MB/s  & 2.41 & 4.2 MB/s   & 164.7 MB/s  & 1.38 \\
\midrule
\multicolumn{7}{l}{\textit{LLM-based baseline}} \\
Conventional AC & 14.10 KB/s & 14.28 KB/s & 0.89 & 17.67 KB/s & 17.35 KB/s & 0.58 \\
\midrule
\multicolumn{7}{l}{\textit{MSMC with Arithmetic Coding}} \\
MSMC + AC ($n{=}1$)     & 81.1 KB/s & 43.2 KB/s & 0.88 & 78.0 KB/s & 39.9 KB/s & 0.58 \\
MSMC + AC ($n{=}2$)     & 86.9 KB/s & 59.9 KB/s & 1.20 & 83.0 KB/s & 56.7 KB/s & 0.97 \\
MSMC + AC ($n{=}4$)     & 87.9 KB/s & 75.9 KB/s & 1.58 & 84.1 KB/s & 72.0 KB/s & \textit{1.48} \\
MSMC + AC ($n{=}6$)     & 86.9 KB/s & 82.7 KB/s & 1.78 & 83.4 KB/s & 79.3 KB/s & \textit{1.76} \\
\midrule
\multicolumn{7}{l}{\textit{MSMC with Rank-Tier Coding}} \\
MSMC + RC ($n{=}1$)     & 100.7 KB/s & 47.6 KB/s & 1.31 & 96.5 KB/s  & 43.8 KB/s & 1.03 \\
MSMC + RC ($n{=}2$)     & 110.5 KB/s & 64.9 KB/s & 1.75 & 104.2 KB/s & 61.3 KB/s & \textit{1.47} \\
MSMC + RC ($n{=}4$)     & 114.2 KB/s & 88.7 KB/s & 2.28 & 107.8 KB/s & 83.8 KB/s & \textit{2.12} \\
MSMC + RC ($n{=}6$)     & 113.7 KB/s & 99.3 KB/s & \textit{2.56} & 107.9 KB/s & 93.5 KB/s & \textit{2.50} \\
\bottomrule
\end{tabular*}
\end{table*}

Table~\ref{tab:main-results} reports all configurations, and Figure~\ref{fig:pareto} plots the two quantities that trade off against each other, bpb and decode throughput.

\paragraph{LLM-based coding compresses far better but runs far slower than classical compressors.}
At $n{=}1$, MSMC\,+\,AC reaches 0.88\,bpb on Wiki and 0.58\,bpb on Code, against 2.41 and 1.38 for xz-9, the strongest classical baseline: the output is 2.7$\times$ and 2.4$\times$ smaller. Decode throughput, however, is 43.2\,KB/s against xz-9's 106.7\,MB/s, more than three orders of magnitude lower. LLM-based compression is therefore attractive only where storage savings outweigh decode cost, as in cold archival data, and the remaining experiments ask how far this throughput gap can be narrowed and at what price.

\paragraph{The pipeline accelerates AC without changing its compression.}
At $n{=}1$ no prediction head is used, so MSMC\,+\,AC and conventional AC produce the same probabilities and differ only in execution. Our pipeline raises encode throughput by 5.75$\times$ on Wiki and 4.42$\times$ on Code, and decode throughput by 3.03$\times$ and 2.30$\times$, while bpb is unchanged up to probablilty quantization (0.89 versus 0.88 on Wiki; 0.58 on Code). Section~\ref{subsec:time} attributes this gain.

\paragraph{The number of tokens per call trades decode throughput for compression.}
Increasing $n$ from 1 to 6 raises MSMC\,+\,AC's decode throughput by 1.91$\times$ on Wiki (43.2$\to$82.7\,KB/s) and 1.99$\times$ on Code (39.9$\to$79.3\,KB/s). The price is a 2.0$\times$ increase in bpb on Wiki (0.88$\to$1.78) and a 3.0$\times$ increase on Code (0.58$\to$1.76). Encode throughput stays nearly flat across $n$ (81.1--87.9\,KB/s on Wiki), because teacher-forced encoding already processes many positions per call and gains nothing from resolving several tokens at once; the extra heads only add projection cost. Within each dataset, every increase in $n$ raises both decode throughput and bpb, and the $\pm1$ standard-deviation ellipses of adjacent operating points do not overlap in bpb (Figure~\ref{fig:pareto}), so the ordering is consistent across files, not an artifact of a single file.

\paragraph{Arithmetic Coding dominates Rank-Tier Coding on the frontier.}
At equal $n$, Rank-Tier Coding decodes 10\% faster than AC at $n{=}1$ and 18--20\% faster at $n{=}6$, but its bpb is 0.43--0.78 higher on Wiki, reflecting its fixed per-tier code lengths (Section~\ref{subsec:rank_tier}). As a result, at every decode throughput that both coders reach, MSMC\,+\,AC attains lower bpb (Figure~\ref{fig:pareto}). Rank-Tier Coding only extends the frontier beyond AC's highest throughput (83\,KB/s on Wiki), and does so at bpb close to or above xz-9.

\paragraph{The useful operating range is bounded by classical compressors.}
Because bpb rises steeply with $n$, high-$n$ configurations lose the advantage that justifies an LLM-based compressor. On Code, MSMC\,+\,AC is worse than xz-9 from $n{=}4$ on, and MSMC\,+\,RC from $n{=}2$ on; on Wiki only MSMC\,+\,RC at $n{=}6$ crosses this line (italic entries in Table~\ref{tab:main-results}). In practice, the useful range is therefore small $n$ with AC: up to $n{=}3$ on Code and somewhat higher on Wiki. Section~\ref{subsec:why-degrade} explains why Code reaches this limit earlier despite starting from lower bpb.

\begin{table}[t]
\centering
\caption{Longer streams: the ten Wiki files concatenated into one 10\,MB input (40\,KB per stream, $B{=}256$). In parentheses: 1\,MB results from Table~\ref{tab:main-results} (4\,KB per stream). Throughput in KB/s.}
\label{tab:long-streams}
\small
\setlength{\tabcolsep}{4pt}
\begin{tabular}{l c c c}
\toprule
Method & Bpb ($\downarrow$) & Encode (KB/s) & Decode (KB/s) \\
\midrule
Conventional AC         & 0.85 (0.89) & 7.7 (14.1)   & 7.8 (14.3) \\
MSMC + AC ($n{=}1$)       & 0.81 (0.88) & 66.0 (81.1)  & 37.7 (43.2) \\
MSMC + AC ($n{=}6$)       & 1.72 (1.78) & 71.0 (86.9)  & 64.0 (82.7) \\
MSMC + RC ($n{=}1$)       & 1.26 (1.31) & 78.4 (100.7) & 40.3 (47.6) \\
MSMC + RC ($n{=}6$)       & 2.52 (2.56) & 86.7 (113.7) & 73.3 (99.3) \\
xz-9                     & 2.17 (2.41) & --           & -- \\
\bottomrule
\end{tabular}
\end{table}

To check that these results do not depend on short streams (about 4\,KB per stream at $B{=}256$), we also compress a 10\,MB concatenation of the Wiki files (Table~\ref{tab:long-streams}). All configurations remain byte-exact, bpb improves slightly for every method including xz, and our pipeline decodes 13--26\% more slowly, which we attribute to longer streams keeping the KV cache between 512 and 1024 positions and thus raising the attention cost per call. The relative conclusions are unchanged: $n{=}6$ still decodes 1.7--1.8$\times$ faster than $n{=}1$, and 3.4--3.7$\times$ faster at $B{=}64$ (not shown), where fewer streams share each call.

\subsection{Where Does the Time Go?}
\label{subsec:time}

\begin{table}[!t]
\centering
\caption{Effect of the coding pipeline (0.5B backbone, FP16, $B{=}256$, $n{=}1$, no prediction heads; mean over 10 files).
Encode/decode throughput in KB/s. Both AC rows use the same model and probabilities; only the execution pipeline differs.
The last row swaps the entropy coder for Rank-Tier coding on the same pipeline.}
\label{tab:pipeline-ablation}
\small
\setlength{\tabcolsep}{3.5pt}
\begin{tabular}{l ccc ccc}
\toprule
& \multicolumn{3}{c}{Wiki} & \multicolumn{3}{c}{Code} \\
\cmidrule(lr){2-4}\cmidrule(lr){5-7}
Configuration & Enc & Dec & Bpb & Enc & Dec & Bpb \\
\midrule
Conventional AC pipeline      & 14.10 & 14.28 & 0.89 & 17.67 & 17.35 & 0.58 \\
Our pipeline, AC              & 81.09 & 43.22 & 0.88 & 78.04 & 39.94 & 0.58 \\
\quad \textit{speedup}        & \textit{5.75$\times$} & \textit{3.03$\times$} & -- & \textit{4.42$\times$} & \textit{2.30$\times$} & -- \\
\midrule
Our pipeline, Rank-Tier       & 100.70 & 47.62 & 1.31 & 96.45 & 43.84 & 1.03 \\
\bottomrule
\end{tabular}
\end{table}

\begin{figure}[t]
\centering
\includegraphics[width=\columnwidth]{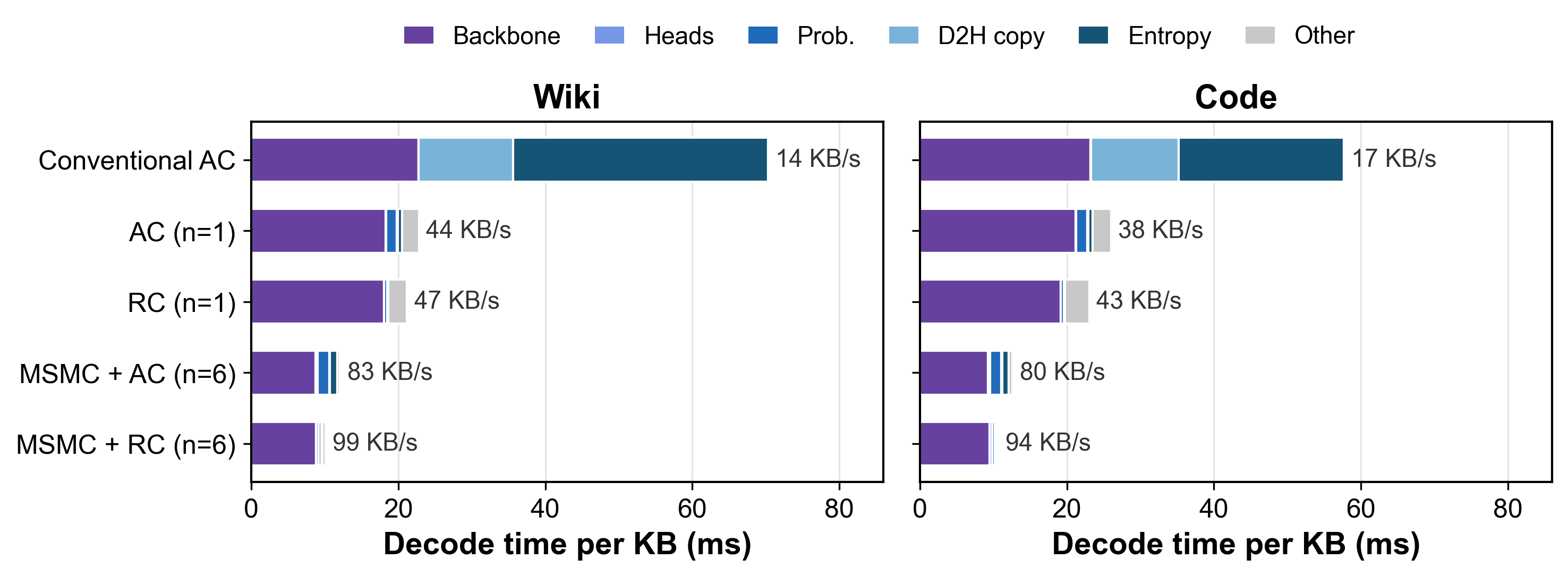}
\caption{Decode time per KB of original data, broken down by stage. Bar length is the un-instrumented decode time. Conventional AC: mean over 10 files; stage times are measured with instrumentation and scaled to the un-instrumented total, so stage shares are approximate. Our pipeline: one representative file per dataset; time not covered by any timed stage is included in ``Other''. The decode throughputs of the representative files are within 5\% of the 10-file means in Table~\ref{tab:main-results}.}
\label{fig:breakdown}
\end{figure}

Table~\ref{tab:pipeline-ablation} separates the effect of the execution pipeline from that of the entropy coder, and Figure~\ref{fig:breakdown} shows which decoding stages account for the difference.

\paragraph{The conventional pipeline is bound by host-side coding, not by the model.}
In conventional AC, CPU-side arithmetic coding and the device-to-host copy of each probability vector together account for 68\% of decode time on Wiki and 60\% on Code (Figure~\ref{fig:breakdown}). Each decoding step must wait for the copy before the coder can run, and the next backbone call must wait for the coder to emit a token, so the GPU idles during most of each step.

\paragraph{Our pipeline makes decoding backbone-bound.}
With probability processing and coding kept on the device and batched across streams, these two stages almost vanish. On the representative files of Figure 5, the copy drops to about 0.1 ms/KB, and entropy coding to 0.59 ms/KB on Wiki and 0.62 ms/KB on Code. The backbone now accounts for 80\% (Wiki) and 82\% (Code) of decode time. The larger encode speedup in Table~\ref{tab:pipeline-ablation} has a separate source: the batch-invariant kernels let the encoder process whole blocks by teacher forcing while still producing exactly the probabilities the token-by-token decoder will see.

\paragraph{A faster entropy coder would add little.}
On our pipeline, entropy coding takes 2.7\% (Wiki) and 2.3\% (Code) of decode time at $n{=}1$ on these files; probability processing, copy, and coding together take 9.7\% and 8.8\%. By Amdahl's law, replacing AC with any faster entropy coder, such as rANS or FSE~\cite{duda2013ans} with the same predictor and precision, could therefore improve decode throughput by at most about 10\%. The last row of Table~\ref{tab:pipeline-ablation} confirms this from the other direction: swapping AC for the much simpler Rank-Tier coder on the same pipeline raises decode throughput by only 10\%, at a bpb cost of 0.43 on Wiki and 0.45 on Code. The dominant speedup comes from the execution pipeline, not from the choice of coder.

\paragraph{Coder cost re-emerges as the backbone share shrinks.}
At $n{=}6$, backbone time per KB on the same files falls by 2.2$\times$ relative to $n{=}1$ (17.27$\to$7.79\,ms/KB on Wiki; 22.33$\to$10.17 on Code). Coder-side work does not shrink, because it scales with the number of tokens rather than with the number of backbone calls: AC's probability processing and coding now take 21\% of decode time on Wiki and 19\% on Code. This is where Rank-Tier Coding's cheaper per-token work becomes visible, making MSMC\,+\,RC 18--20\% faster than MSMC\,+\,AC at $n{=}6$ (Table~\ref{tab:main-results}). Coder efficiency thus matters more the more aggressively backbone calls are reduced.

\subsection{Why Does Compression Degrade with $n$?}
\label{subsec:why-degrade}

\begin{figure}[t]
\centering
\includegraphics[width=\columnwidth]{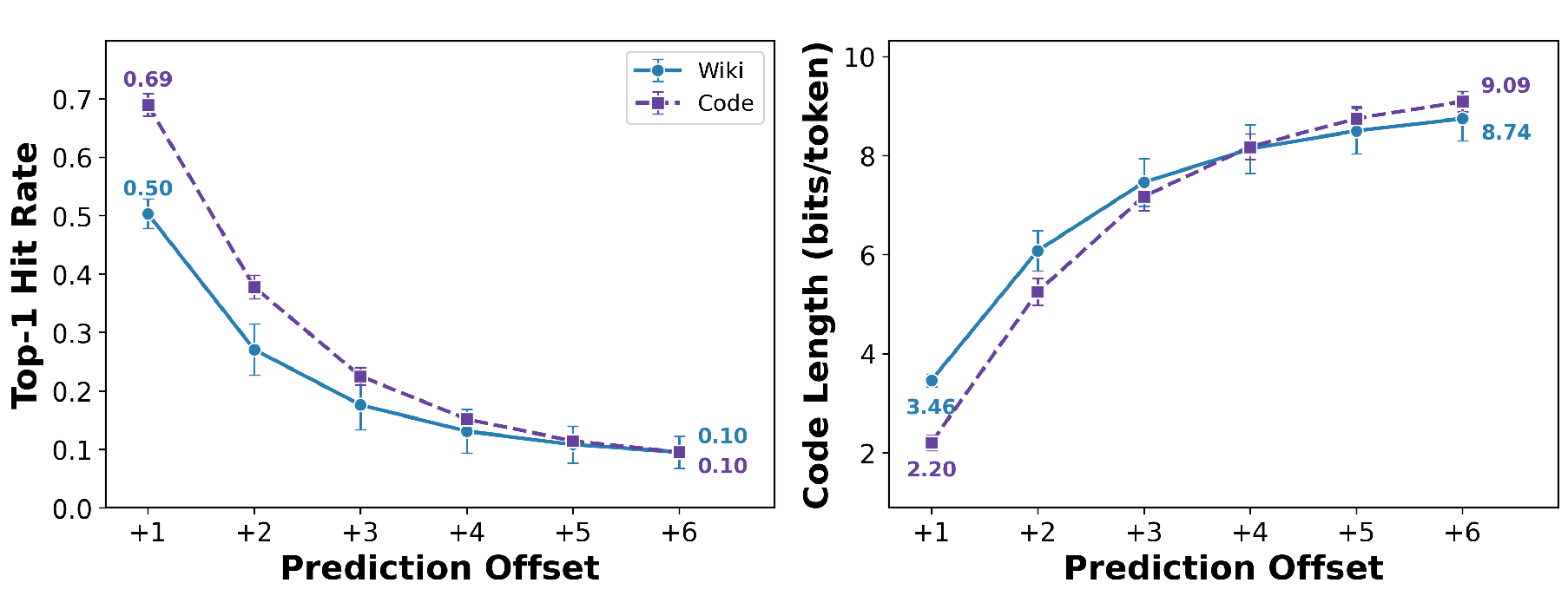}
\caption{Per-offset prediction quality of the domain heads: top-1 hit rate (left) and average code length in bits per token (right) at each prediction offset. Offset $+1$ is the frozen backbone LM head. Error bars: $\pm1$ standard deviation across the 10 files of each dataset.}
\label{fig:head-accuracy}
\end{figure}

With $n$ tokens resolved per call, the token at offset $+k$ is predicted from a hidden state that is $k$ positions behind it. Figure~\ref{fig:head-accuracy} measures how prediction quality falls with distance. This decline has two sources that the figure does not separate: the information lost by not conditioning on the $k-1$ intervening tokens, which no head can recover (Section~\ref{subsec:price}), and the prediction error of our trained heads, which better-trained heads could reduce. The figure therefore reflects the heads we trained, not a lower bound on what multi-token coding must pay.

\paragraph{Prediction quality decays quickly with distance.}
Top-1 hit rate falls from 0.50 at offset $+1$ to 0.10 at $+6$ on Wiki, and from 0.69 to 0.10 on Code. Correspondingly, the average code length grows from 3.46 to 8.74 bits/token on Wiki and from 2.20 to 9.09 bits/token on Code. Most of the loss occurs at the first head ($+2$); later heads add progressively less.

\paragraph{Per-offset code lengths account for the observed bpb.}
Each decoding round of size $n$ codes one token at each offset $+1,\dots,+n$, so the expected code length per token is the mean of the first $n$ per-offset code lengths. Dividing this mean by the average number of bytes per token reproduces the measured bpb of Table~\ref{tab:main-results}. The compression--throughput trade-off of Section~\ref{subsec:overall} is thus fully explained by how well the heads predict at each distance, not by any loss in the coding mechanism itself.

\paragraph{Why Code degrades faster.}
Code is more predictable than Wiki one token ahead (hit rate 0.69 versus 0.50), which gives it its lower bpb at $n{=}1$. From offset $+4$ on, however, its code length matches or exceeds Wiki's. Locally, code is highly regular, but which identifier or statement comes several tokens ahead is hard to predict. This is why Code's bpb rises 3.0$\times$ from $n{=}1$ to $n{=}6$ against Wiki's 2.0$\times$, and why Code crosses the xz-9 line at a smaller $n$. Better far-offset prediction, for example from larger heads or longer head training, is therefore the main lever for widening MSMC's useful range.

\subsection{When Does Multi-Token Decoding Pay Off?}
\label{subsec:when-pays}

\begin{figure}[t]
\centering
\includegraphics[width=\columnwidth]{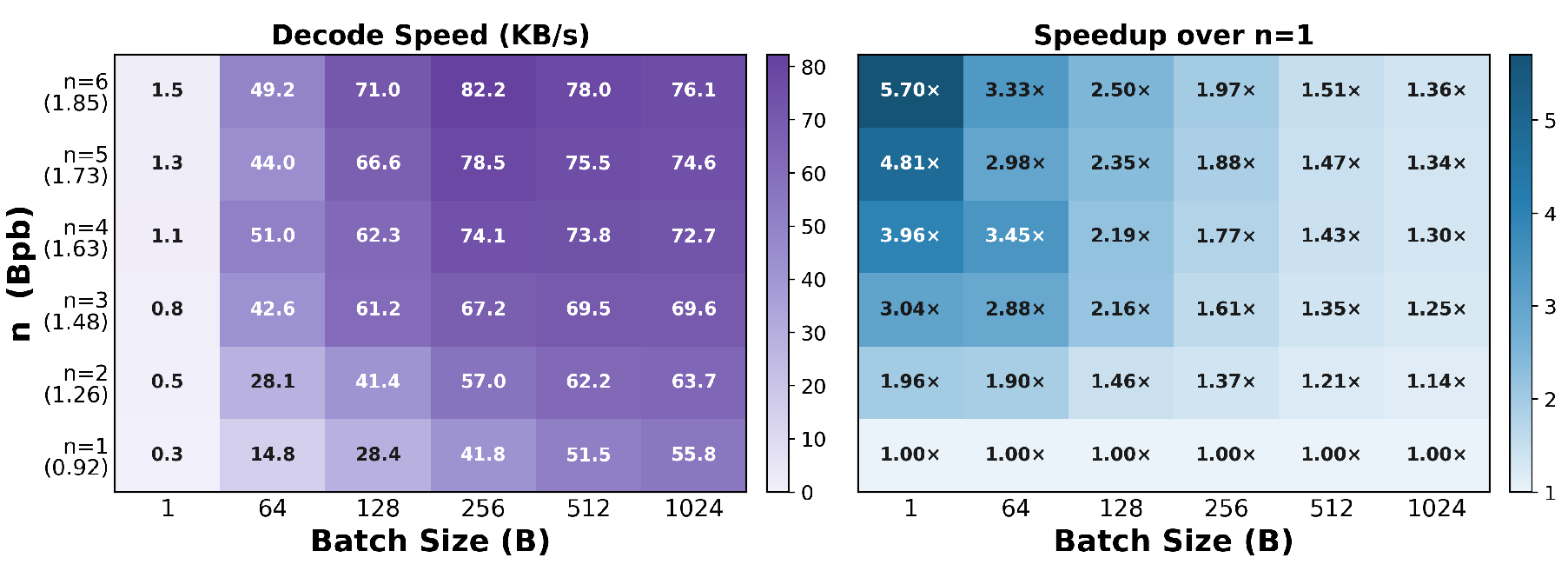}
\caption{Decode throughput of MSMC\,+\,AC across batch size $B$ and tokens per call $n$ (left), and speedup over $n{=}1$ at the same $B$ (right), on one representative Wiki file. Row labels give the bpb of each $n$ on that file.}
\label{fig:bxn}
\end{figure}

MSMC reduces the number of sequential backbone calls; batching raises the work done per call. Both aim to use the GPU better, so their benefits should interact. Figure~\ref{fig:bxn} sweeps both.

\paragraph{At small batch sizes, the speedup approaches $n$.}
At $B{=}1$, resolving $n$ tokens per call speeds up decoding by 1.96$\times$, 3.04$\times$, 3.96$\times$, 4.81$\times$, and 5.70$\times$ for $n{=}2$ to 6, close to the ideal speedup of $n$ derived in Section~\ref{sec:msmc}. With a single stream, each backbone call is latency-bound: its cost is dominated by reading the weights and launching kernels, so processing a few more positions per call is almost free.

\paragraph{At large batch sizes, the two techniques compete.}
As $B$ grows, each call becomes compute-bound and extra positions cost proportionally more. The speedup of $n{=}6$ falls to 1.97$\times$ at $B{=}256$ and 1.36$\times$ at $B{=}1024$. Absolute throughput behaves accordingly: $n{=}1$ keeps improving up to $B{=}1024$ (55.8\,KB/s), while $n{=}6$ peaks at $B{=}256$ (82.2\,KB/s) and declines beyond it. The highest throughput in the sweep is still reached with $n{=}6$.

\paragraph{Implication.}
Multi-token decoding is most valuable when the number of independent streams is limited: decompressing a single small block, serving random access into an archive, or restoring data under a latency constraint. In these settings, it delivers speedups that additional batching cannot. In throughput-oriented bulk decompression with large batches, its benefit is smaller and must be weighed against the bpb cost quantified in Section~\ref{subsec:why-degrade}. We note that absolute single-stream throughput remains low (1.5\,KB/s at $B{=}1$, $n{=}6$), which bounds the latency-sensitive use cases LLM-based compression can serve at this model scale.

\subsection{Does a Separate Draft Model Help?}
\label{subsec:sdmc-vs-msmc}

Section~\ref{sec:speculative} explored an alternative way to reduce expensive backbone calls: a small draft model (0.5B) proposes tokens and a large target model (7B) is invoked only when needed. We evaluate this design against single-model coding on the same 7B target. SDMC requires FP32 for lossless decoding in our implementation (Section~\ref{subsec:sdmc_limits}), so all methods in this subsection run in FP32 with matched batching (encode $B{=}128$, decode $B{=}1$, the largest settings SDMC supports on one H100). Because 7B-scale runs are expensive, this comparison uses one representative file per dataset and should be read as qualitative.

\begin{table}[t]
\centering
\caption{SDMC versus single-model coding on the 7B backbone under matched FP32 precision and batching (encode $B{=}128$, decode $B{=}1$; one representative file per dataset). SDMC uses the un-fine-tuned 0.5B base model as its draft model. Throughput in KB/s.}
\label{tab:sdmc}
\small
\setlength{\tabcolsep}{3.5pt}
\begin{tabular}{l cc cc cc}
\toprule
& \multicolumn{2}{c}{Bpb ($\downarrow$)} & \multicolumn{2}{c}{Encode} & \multicolumn{2}{c}{Decode} \\
\cmidrule(lr){2-3}\cmidrule(lr){4-5}\cmidrule(lr){6-7}
Method & Wiki & Code & Wiki & Code & Wiki & Code \\
\midrule
AC (7B)             & 0.65 & 0.48 & 4.05  & 3.68  & 0.28 & 0.25 \\
RC (7B)             & 1.02 & 0.88 & 16.76 & 14.49 & 0.24 & 0.23 \\
SDMC (0.5B\,+\,7B)  & 0.97 & 0.70 & 8.58  & 7.81  & 0.29 & 0.28 \\
\bottomrule
\end{tabular}
\end{table}

\paragraph{SDMC's gain is limited to decoding and is small.}
Relative to single-model Rank-Tier Coding on the same target, SDMC decodes about 20\% faster (0.29 versus 0.24\,KB/s on Wiki; 0.28 versus 0.23 on Code), because tokens the draft model predicts correctly skip the 7B call. Encoding, however, is about 2$\times$ slower (8.58 versus 16.76\,KB/s on Wiki; 7.81 versus 14.49 on Code): teacher-forced encoding already batches positions for the single-model coder, so the draft model only adds a second, sequential model to the loop. For comparison, MSMC obtains a 1.96$\times$ single-stream speedup already at $n{=}2$ (Section~\ref{subsec:when-pays}), with one model and no extra encode cost; that measurement uses the 0.5B backbone in FP16, so the comparison is indicative rather than matched.

\paragraph{Compression lies between RC and AC}
SDMC reaches 0.97 and 0.70\,bpb, below single-model RC (1.02 and 0.88) but well above AC on the same 7B model (0.65 and 0.48). The gap to AC comes from SDMC’s fixed per-token code (Section~\ref{SDMC_design}), not from its decoding schedule. Fine-tuning the draft model on the target file raises its top-1 hit rate (from 47.7\% to 55.3\% on Wiki with standard cross-entropy, and to 56.5\% with the additional margin term) and lowers bpb accordingly, but does not change the encode-side cost.
\section{Related Work}
\label{sec:related}

\paragraph{Classical and learned compression.}
Dictionary compressors such as gzip~\cite{deutsch1996gzip, ziv1977lz77}, zstd~\cite{collet2021zstd}, and FSST~\cite{boncz2020fsst} decode at hundreds of MB/s but exploit mostly local, literal redundancy. Statistical compressors pair a predictive model with arithmetic coding~\cite{rissanen1976generalized, witten1987arithmetic} or ANS~\cite{duda2013ans}, from PPM~\cite{cleary1984ppm, bell1990text} and context mixing~\cite{mahoney2005pzip, knoll_cmix} to neural predictors~\cite{goyal2019deepzip, bellard2019nncp}. L3TC~\cite{l3tc} reaches MB/s-level decoding with a small RWKV~\cite{rwkv} model at a lower compression rate; it makes each call cheaper by shrinking the model, which is complementary to our study with a fixed model. Database systems favor lightweight schemes that decode fast enough to integrate with query execution~\cite{abadi2006integrating, damme2017lightweight, afroozeh2023fastlanes, afroozeh2024alp}, including on GPUs~\cite{shanbhag2022tile}, and learned models have been used for semantic compression of tabular data~\cite{ilkhechi2020deepsqueeze}.

\paragraph{LLM-based compression.}
LLMZip~\cite{valmeekam2023llmzip} pairs LLaMA-7B~\cite{touvron2023llama} with arithmetic coding or with a compressed stream of token ranks, FineZip~\cite{finezip} adds online fine-tuning and batching, Llamazip~\cite{llamazip} uses the same mechanism to detect training data, and Mao et al.~\cite{ntp2025} compress LLM-generated text. These works target compression rate and report decoding throughput briefly, if at all.

\paragraph{Throughput of LLM-based compression.}
Kipf et al.~\cite{waiting2026cidr} benchmark LLM-based compressors and conclude that, at current prices, their compute cost takes about a decade to amortize, leaving cold archival data as the most plausible use; their public implementation is our baseline. Adams et al.~\cite{adams2026highthroughput} find that, once inference is optimized, arithmetic coding limits compression speed, and they transfer only each target token's interval to the host while pipelining coding with inference. We apply the same principles, and additionally make block-parallel encoding lossless in half precision and remove the corresponding redundancy from decoding, where the target interval is not known in advance. We then study the sequential dependency that remains.

\paragraph{Rank-based coding.}
Coding a symbol by its rank goes back to Shannon~\cite{shannon1951prediction}. With LLM predictors, rank streams have been passed to general-purpose compressors~\cite{valmeekam2023llmzip, finezip}, thresholded with an escape~\cite{rankthreshold2025}, and bit-packed or Huffman-coded~\cite{adams2026highthroughput}.

\paragraph{Numerical determinism.}
He et al.~\cite{he2025nondeterminism} trace the non-reproducibility of LLM inference to kernels that lack batch invariance, and design kernels whose per-row results do not depend on the rest of the batch. Lossless compression needs more than run-to-run reproducibility, since encoder and decoder invoke the model with different shapes. We use such kernels to make the two agree in half precision, and report a case, dual-model coding, in which ours do not suffice.

\paragraph{Acceleration techniques from generation.}
Speculative decoding~\cite{leviathan2023speculative, chen2023speculative, xia2024speculative} and EAGLE~\cite{li2024eagle} let a draft model propose tokens that a target model verifies. Blockwise parallel decoding~\cite{stern2018blockwise}, Medusa~\cite{cai2024medusa}, and multi-token prediction~\cite{gloeckle2024multitoken} predict several future tokens from one hidden state, and Hydra~\cite{ankner2024hydra} conditions later heads on earlier drafts. Compression changes the setting in three ways. The encoder knows the true sequence, so drafts never need verification (SDMC). A poor prediction at a later offset costs bits, not only time, which gives rise to the overhead $\Delta_n$ that MSMC pays. And encoder and decoder must produce bit-identical predictions.
\section{Conclusion}
\label{sec:discussion}

We separated the decoding time of LLM-based lossless compression into execution redundancy and sequential dependency. Removing execution redundancy with batch-invariant kernels and a GPU-resident pipeline leaves model invocation as the main decompression bottleneck: entropy coding accounts for little of the remaining time, and multi-token prediction can trade fewer calls for more bits only at a rate set by how well its predictors work. This opens research into models designed for compression rather than generation, with cheaper inference per token and multi-token predictions that condition on more of the tokens they resolve together.


\bibliographystyle{ACM-Reference-Format}
\bibliography{ref}

\end{document}